\documentclass[epjCONF,columns]{svjour} 
\usepackage{graphics}
\usepackage{xspace}
\usepackage{amsmath}

\usepackage[varg]{txfonts} 
\usepackage[latin1]{inputenc}

\newcommand{\GeV}{\ensuremath{\,\text{Ge\hspace{-.08em}V}}\xspace}

\newcommand{\TeV}{\ensuremath{\,\text{Te\hspace{-.08em}V}}\xspace}

\newcommand{\fbinv} {\mbox{\ensuremath{\,\text{fb}^\text{$-$1}}}\xspace}
\newcommand{\pbinv} {\mbox{\ensuremath{\,\text{pb}^\text{$-$1}}}\xspace}

\newcommand{\pt}{\ensuremath{p_{\mathrm{T}}}\xspace}

\newcommand {\etal}{\mbox{et al.}\xspace} 

\newcommand{\ZPSSM}{\ensuremath{Z'_\text{SSM}}\xspace}
\newcommand{\ZPPSI}{\ensuremath{Z'_\psi}\xspace}
\newcommand{\GKK}{\ensuremath{\mathrm{G}_\text{KK}}\xspace}

\newcommand{\Mgg}{\ensuremath{{M_{\gamma\gamma}}}\xspace}
\newcommand{\ktilde}{\ensuremath{{\tilde{k}}}\xspace}

\newcommand{\mm}{\ensuremath{{\mu^{+}\mu^{-}}}\xspace}
\newcommand{\ee}{\ensuremath{{ee}}\xspace}
\newcommand{\ttbar}{\ensuremath{{t\overline{t}}}\xspace}

\session-title{2011 Hadron Collider Physics symposium (HCP-2011)}

\begin{document}
\title{Searches for high mass resonances with the CMS detector}
\author{Toyoko J. Orimoto\inst{1}\fnmsep\thanks{\email{toyoko.orimoto@cern.ch}}, on behalf of the CMS Collaboration}
\institute{$^1$ CERN, CH-1211 Geneva 23, Switzerland}
\abstract{
New heavy resonances are predicted by many extensions of the standard model of particle physics.
Recent results for high mass resonance searches with the Compact Muon Solenoid detector, in the diphoton, dilepton, dijet and \ttbar channels,  are discussed.  Limits for numerous benchmark models are presented.
} 
\maketitle

\section{Introduction}

We present recent results for high mass resonance searches with the Compact Muon Solenoid (CMS) detector. The searches are conducted in a model independent manner, looking for excesses in the diphoton, dilepton, dijet, and \ttbar invariant mass spectra.  As no excesses above standard model (SM) expectations are observed, limits  are computed, probing a variety of benchmark models, such as those predicting Randall-Sundrum (RS) gravitons  \cite{ref_rsg}, extra heavy gauge bosons ($Z'$, $W'$) \cite{ref_gauge}, and other exotic phenomena.

\section{The CMS Detector}\label{sec:cms}

The central feature of the CMS apparatus is a superconducting solenoid, of 6m internal diameter, providing a field of 3.8~T. Within the field volume are the silicon pixel and strip tracker, the crystal electromagnetic calorimeter (ECAL) and the brass and scintillator hadron calorimeter (HCAL). Muons are measured in gas-ionization detectors embedded in the steel return yoke. In addition to the barrel and endcap detectors, CMS has extensive forward calorimetry.  
A more detailed description can be found in Ref.~\cite{JINST}.

\section{Diphoton Resonances}\label{sec:diphoton}

\begin{figure}
\resizebox{0.9\columnwidth}{!}{%
  \includegraphics{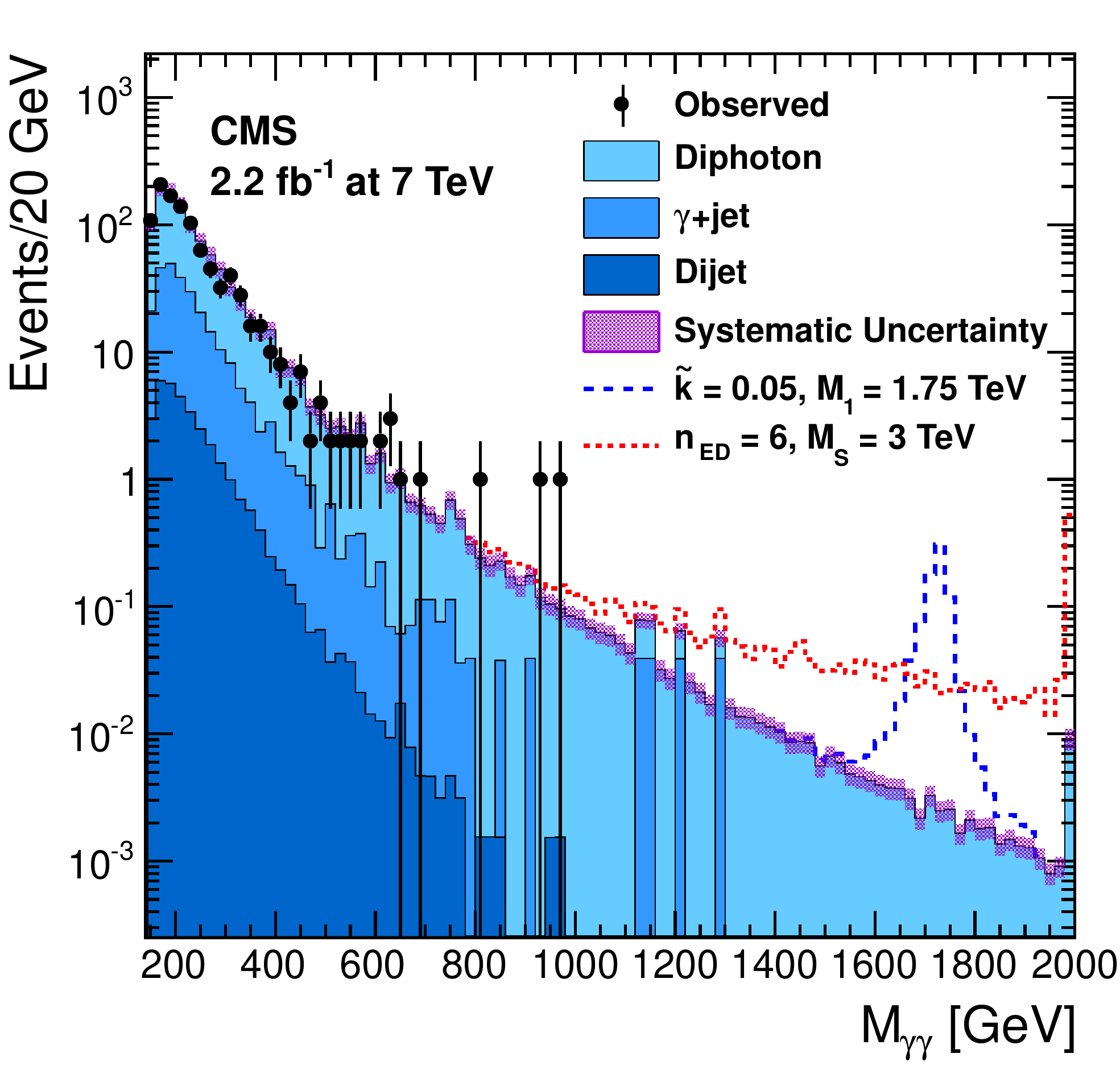} }
\caption{Diphoton invariant mass spectrum.  Observed event yields (points with error bars) and background expectations (filled solid histograms) as a function of the diphoton invariant mass. The shaded band around the background estimation corresponds to the average systematic uncertainty over the spectrum. The last bin includes the sum of all contributions for $\Mgg > 2.0$ \TeV. The simulated distributions for two signal hypotheses are shown for comparison as dotted (ADD) and dashed (RS) lines.}
\label{fig:mass_diphoton}      
\end{figure}

\begin{figure}
\resizebox{\columnwidth}{!}{%
  \includegraphics{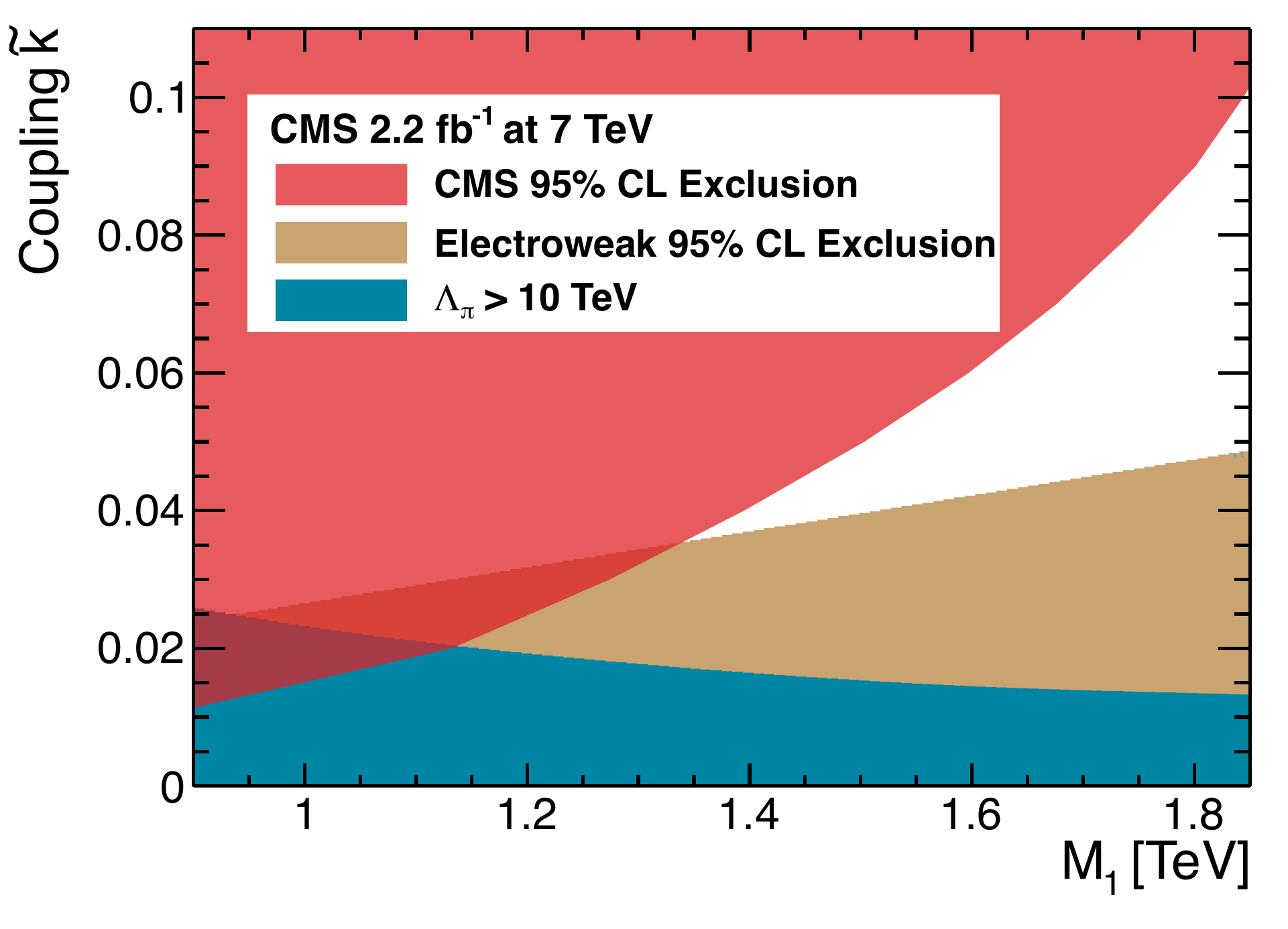} }
\caption{
The $95\%$ CL exclusion region for the RS1 graviton model in the $M_{1}$-$\ktilde$ plane.  The expected limits coincide very closely with the measured limits and so are not shown in the figure. Also shown are bounds due to electroweak constraints and naturalness ($\Lambda_{\pi} > 10 \TeV$). Perturbativity requirements bound $\ktilde < 0.10$.
} 
\label{fig:limit_diphoton}      
\end{figure}

Kaluza-Klein (KK) gravitons predicted by RS warped extra dimensions may manifest themselves as high mass resonances in the diphoton invariant mass spectrum.  The diphoton channel has the advantage that the branching ratio for spin-2 gravitons is twice that to leptons.  
Two isolated photons are selected with $E_T$ $>$ 70 \GeV and $|\eta|$ $<$ 1.44.   The resulting diphoton invariant mass distribution with 2.2\fbinv of data is shown in Fig.~\ref{fig:mass_diphoton}~\cite{diphoton}.  The expected background arising from irreducible SM diphoton production is estimated using simulation, scaled by a next-to-leading order mass dependent $K$ factor.  
Instrumental backgrounds, arising from $\gamma$+jet and dijet processes, in which the jets are misidentified as photons, are estimated using a data-driven fake rate  method.  
Observing no excess in the diphoton invariant mass distribution above SM expectations, upper limits are set on the production cross section for RS gravitons, using the CL$_S$ technique~\cite{CLs1,CLs2}.  The limits on the cross section are translated into lower limits on the model parameters (Fig.~\ref{fig:limit_diphoton}), where M$_1$ is the mass of the first graviton excitation, and \ktilde is a dimensionless parameter which quantifies the strength of the graviton coupling to SM fields.
We exclude at the $95\%$ confidence level (CL) resonant graviton production in the RS1 model with values of $\text{M}_1$ $<$ $0.86-1.84$ \TeV, depending on \ktilde.


%

\section{Dilepton Resonances}\label{sec:dilepton}

\begin{figure}
\resizebox{0.88\columnwidth}{!}{%
  \includegraphics{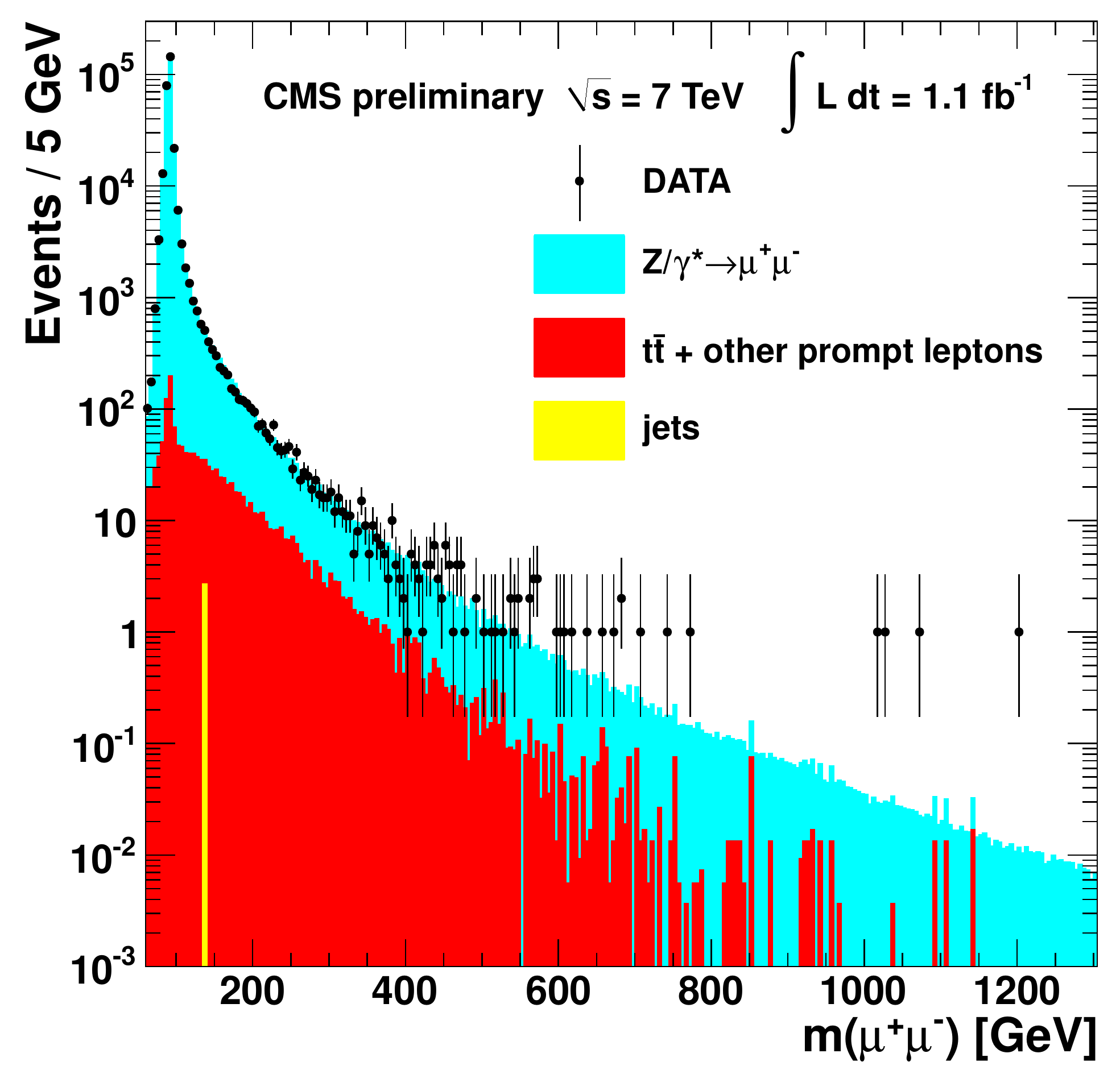} }
  \resizebox{0.88\columnwidth}{!}{%
    \includegraphics{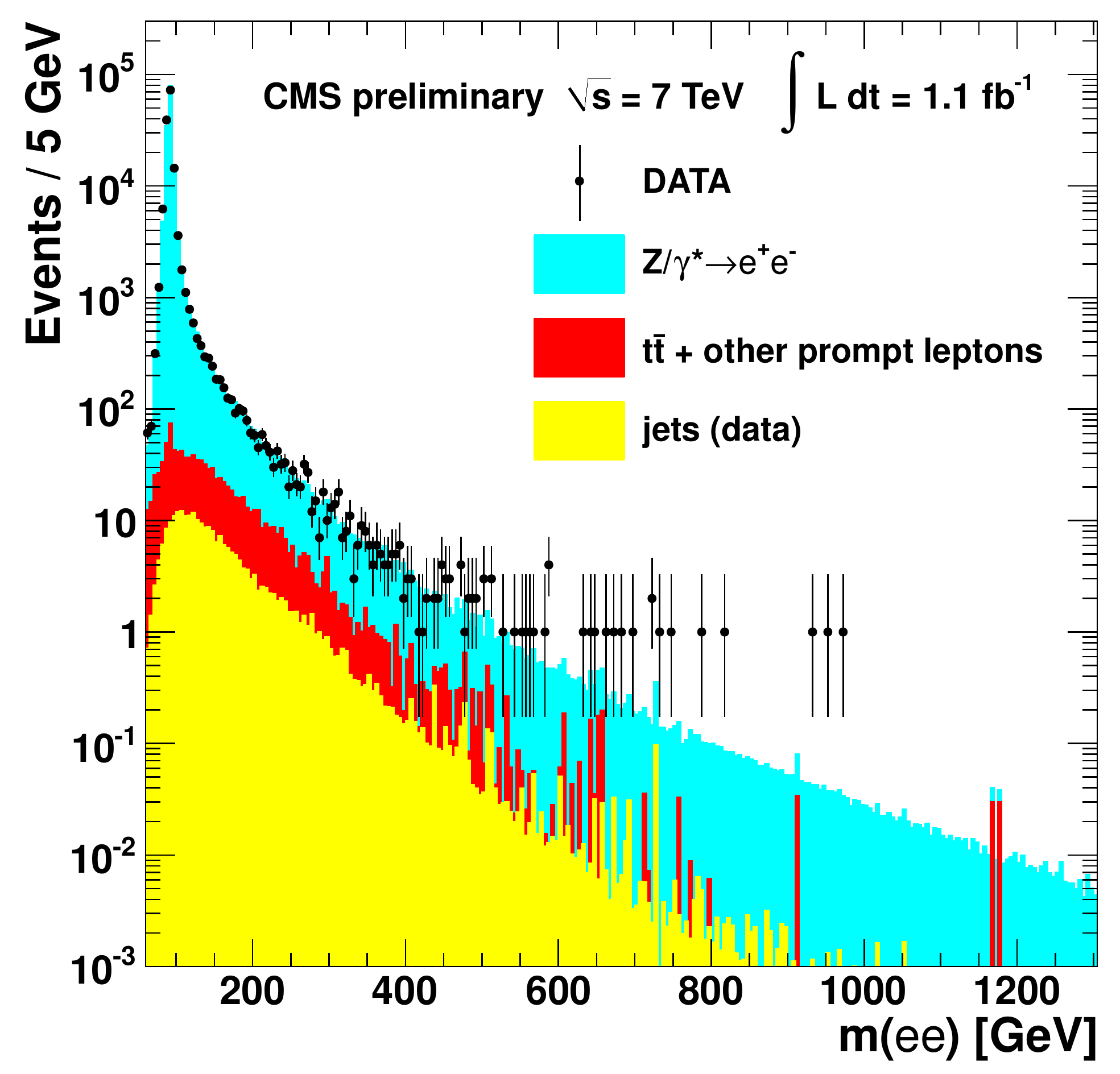} }
\caption{
Invariant mass spectrum of \mm (top) and \ee (bottom) events. Data is represented by points with error bars. The uncertainties on the data points (statistical only) represent $68\%$ confidence intervals for the Poisson means.  The filled histograms represent the expectations from SM processes.
}
\label{fig:mass_dilepton}      
\end{figure}

High mass dilepton resonances may arise in models with extra heavy gauge bosons, as well as in the RS warped extra dimension scenario.  
In this search, two isolated leptons with $\pt >$ 35 \GeV (40 \GeV for endcap electrons) are required, in addition to further quality criteria. In the case of dimuons, opposite-sign is required.  The resulting dielectron and dimuon invariant mass distributions from 1.1\fbinv of data can be seen in Fig.~\ref{fig:mass_dilepton} \cite{dilepton}.  
The expected background from irreducible SM Drell-Yan production is estimated using simulation, normalized to the $Z^0$ peak in data.  Backgrounds with prompt leptons (\ttbar, $tW$, dibosons) are cross-checked using a data-driven method counting $e$-$\mu$ pairs.  Backgrounds with jets misidentified as leptons ($W$+jet, dijets) are estimated using a fake rate measured from a jet-enriched data sample.  Lastly, cosmic muon backgrounds are rejected by imposing topological criteria.  

\begin{figure}
\resizebox{\columnwidth}{!}{%
  \includegraphics{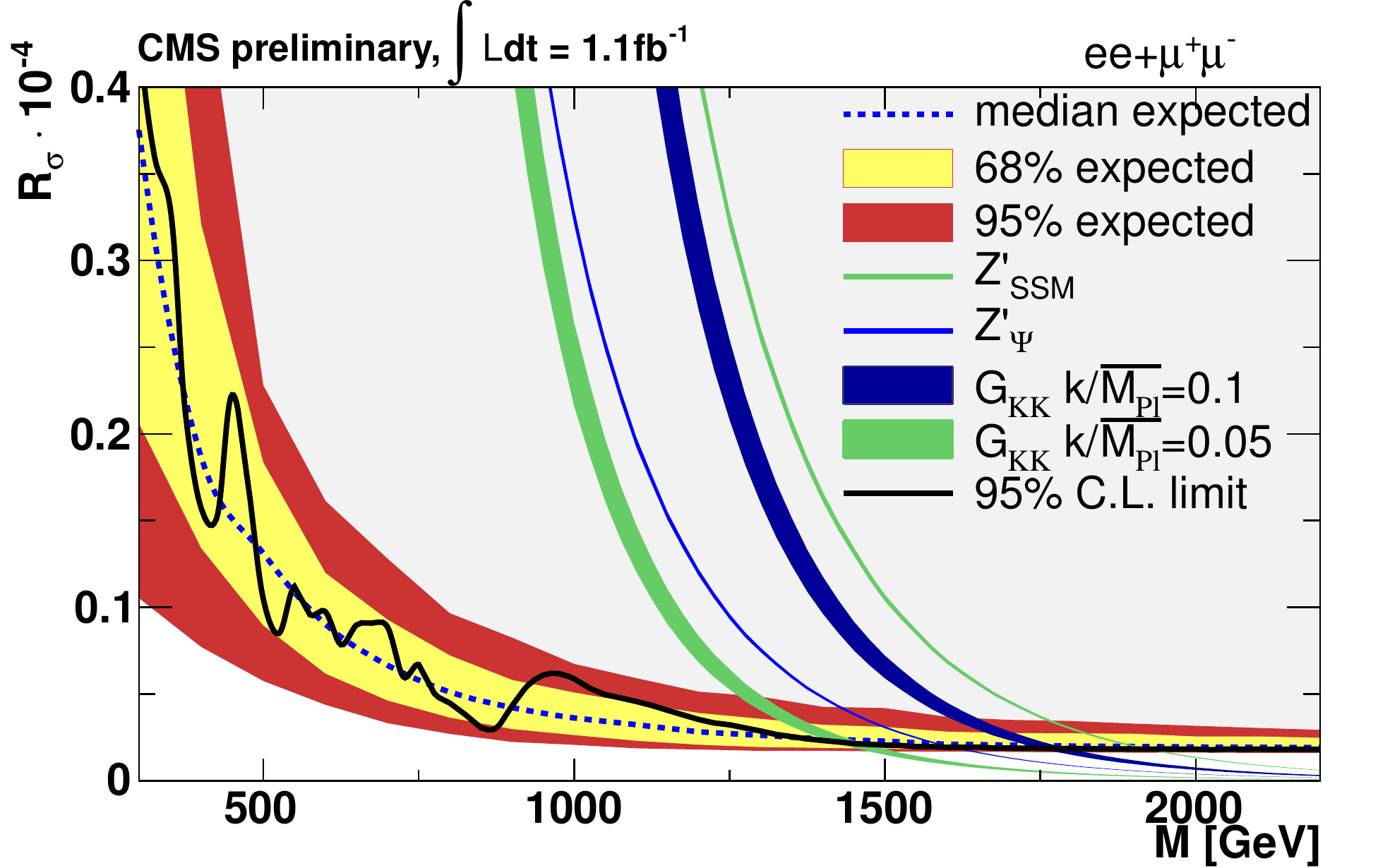} }
\caption{
Upper limits on the production ratio R$_\sigma$ of cross
section times branching fraction into lepton pairs as a function of resonance mass, for \ZPSSM, \ZPPSI, and \GKK. The limits are shown from the combined dilepton (\ee + \mm) result. Shaded yellow and red bands correspond to the $68\%$ and $95\%$ quantiles for the expected limits. The predicted cross section ratios are shown in bands, with widths indicating theoretical uncertainties.
} 
\label{fig:limit_dilepton}      
\end{figure}

The dilepton analysis is a shape-based search, making no assumptions on the absolute background rate; this is achieved by normalizing the results to the $Z^0$ peak.  
We set limits, using a Bayesian technique, on the ratio (R$_{\sigma}$)  of the cross section for $Z'$ (or \GKK) production to the cross section for SM $Z^{0}$ production (Fig.~\ref{fig:limit_dilepton}).  The limits on R$_{\sigma}$ can be interpreted as lower limits on $Z'$ ($\GKK$) mass. We exclude at $95\%$ CL a $Z'$ with SM-like couplings (\ZPSSM) with mass $< 1940 \GeV$, the superstring-inspired \ZPPSI $<$ 1620 \GeV, and RS \GKK $< 1450 ~(1780) \GeV$ for \ktilde=0.05 (0.10).


\section{Dijet Resonances}\label{sec:dijet}

Dijets can be used to probe a variety of beyond-the-SM signatures, including string resonances~\cite{ref_string}, $\mbox{E}_6$ diquarks~\cite{ref_diquark}, excited quarks~\cite{ref_qstar}, axigluons~\cite{ref_axi}, colorons~\cite{ref_coloron}, $\mbox{W}^{\prime}$ and $\mbox{Z}^{\prime}$, and RS gravitons.  
We perform a general model independent shape-based search for three types of resonances (qq, qg, gg), the differences arising from final state radiation (FSR).

\begin{figure}
\resizebox{0.9\columnwidth}{!}{%
  \includegraphics{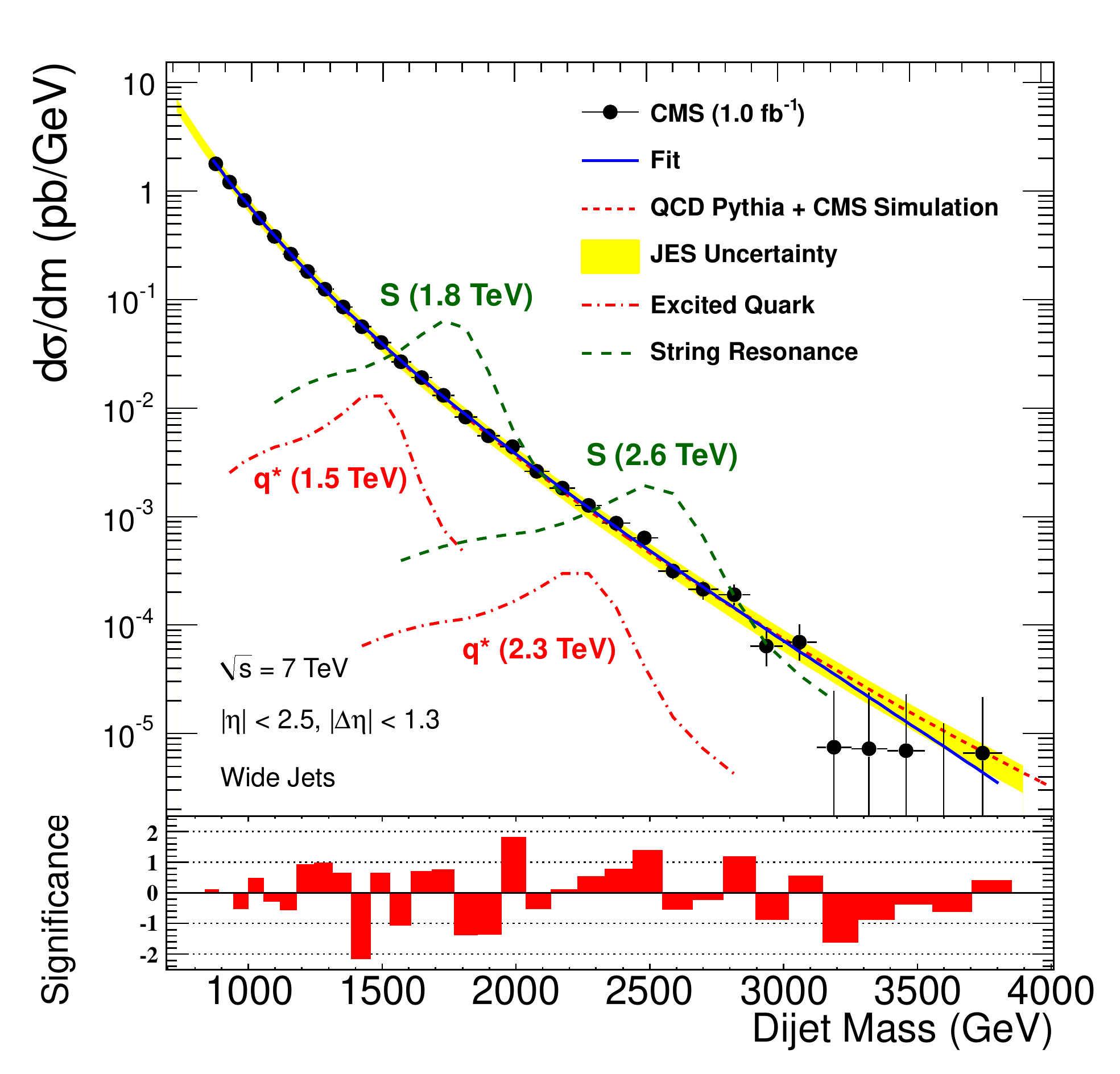} }
\caption{
    Dijet mass spectrum from wide jets (points) compared to a smooth fit (solid) and to predictions including detector simulation of QCD (short-dashed), excited quark signals (dot-dashed),  and string resonance signals (long-dashed). The QCD prediction has been normalized to the data. The error bars are statistical only.  The shaded band shows the systematic uncertainty in the jet energy scale (JES).   The bin-by-bin significance of the data-fit difference is shown at bottom.
}
\label{fig:mass_dijet}      
\end{figure}

\begin{figure}
\resizebox{\columnwidth}{!}{%
  \includegraphics{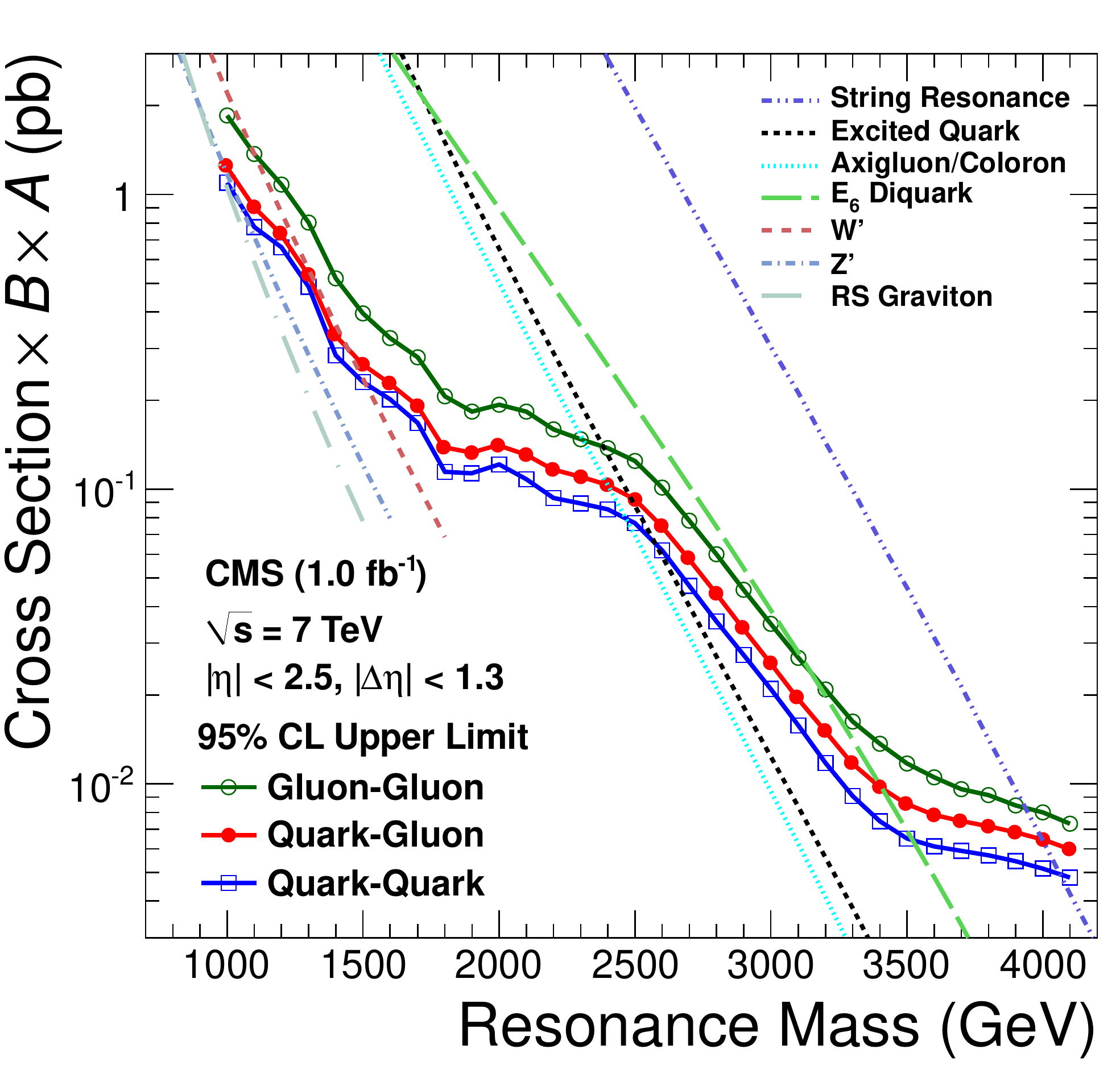} }
    \caption{The 95\% CL upper limits on cross section ($\sigma$) and branching ratio ($B$) and acceptance ($A$),  for dijet resonances of type gluon-gluon (open circles), quark-gluon (solid circles), and quark-quark (open boxes),
compared to theoretical predictions for string resonances, $\mbox{E}_6$ diquarks, excited quarks, axigluons, colorons, new gauge bosons $\mbox{W}^{\prime}$ and $\mbox{Z}^{\prime}$, and RS gravitons.}
\label{fig:limit_dijet}      
\end{figure}

We require the two leading jets have $|\eta|$ $<$2.5 and $|\Delta\eta|$ $<$1.3 and dijet invariant mass $>$ 838\GeV.  
To recover radiation lost through FSR and to improve the dijet mass resolution, we combine particle flow \cite{pflow} jets with the anti-k$_T$ algorithm ($R=0.5$) into ``wide jets".
QCD multijets comprise the main background, following a smoothly falling dijet mass distribution predicted by the SM.
Fig.~\ref{fig:mass_dijet} shows the dijet invariant mass spectrum with 1.0\fbinv, where the expected background from QCD multijets is described with a functional fit \cite{dijet}.  The systematic uncertainties from the jet energy scale and resolution are  $2\%$ and $10\%$, respectively.
The $95\%$ CL upper limits on the product of the cross section and branching ratio and acceptance, computed using a Bayesian approach, are shown in Fig.~\ref{fig:limit_dijet}.  We exclude masses for string resonances $<$ 4.00\TeV, $E_6$ diquarks $<$ 3.52\TeV, excited quarks $<$ 2.49\TeV, axigluons/ colorons $<$ 2.47\TeV, and $W'$ bosons $<$ 1.51\TeV.  No limits on $Z'$ or RS gravitons are set.
The use of wide-jets improves the limits by $20\%$ for gg, $10\%$ for qg, and $5\%$ for qq. 


\section{$t\bar{t}$ Resonances: Semileptonic Decay}\label{sec:ttsemileptonic}

New bosons with enhanced coupling to the top quark appear in many SM extensions, such as those predicting axigluons and KK gluons~\cite{rs_gluon_1}.
We present a search for heavy \ttbar resonances in the semileptonic $(qqb)(\mu\nu b)$ final state, focusing on highly boosted top pairs with decay products narrowly collimated along the direction of the top.
Backgrounds arise from SM \ttbar, $W/Z+$jets, single top, and QCD multijets.

For high mass \ttbar, the decay products of the hadronic-decaying top can have small opening angles in the detector frame. Thus, instead of requiring four jets, we require two particle flow jets with $\pt$ $>$ 50\GeV and $|\eta|$ $<$ 2.4, with the leading jet $\pt$ $>$ 250\GeV; jets are reconstructed with the anti-k$_T$ algorithm ($R=$0.5).
The high top \pt also results in low $\Delta R = \sqrt{(\Delta \phi)^{2} + (\Delta \eta)^{2}}$ between the $\mu$ and $b$, making it difficult to require the muon be well-isolated.  To suppress QCD multijet backgrounds, we thus apply a two-dimensional requirement,  $\Delta R >$ 0.5 or $p_{\text{T,rel}}$ $>$ 25\GeV, where $p_{\text{T,rel}}$ is the magnitude of the $p_\mu$ component orthogonal to the jet axis.
In addition, muons are required to have $\pt$ $>$ 35\GeV and $|\eta|$ $<$ 2.1.  
Events with additional muons or electrons (from \ttbar and $Z^0$ decays) are vetoed.  Lastly, $H_{\text{T,lep}}$, the scalar sum of the muon \pt and missing transverse energy (MET), is required to be $>$ 150 \GeV.

\begin{figure}
\resizebox{\columnwidth}{!}{%
  \includegraphics{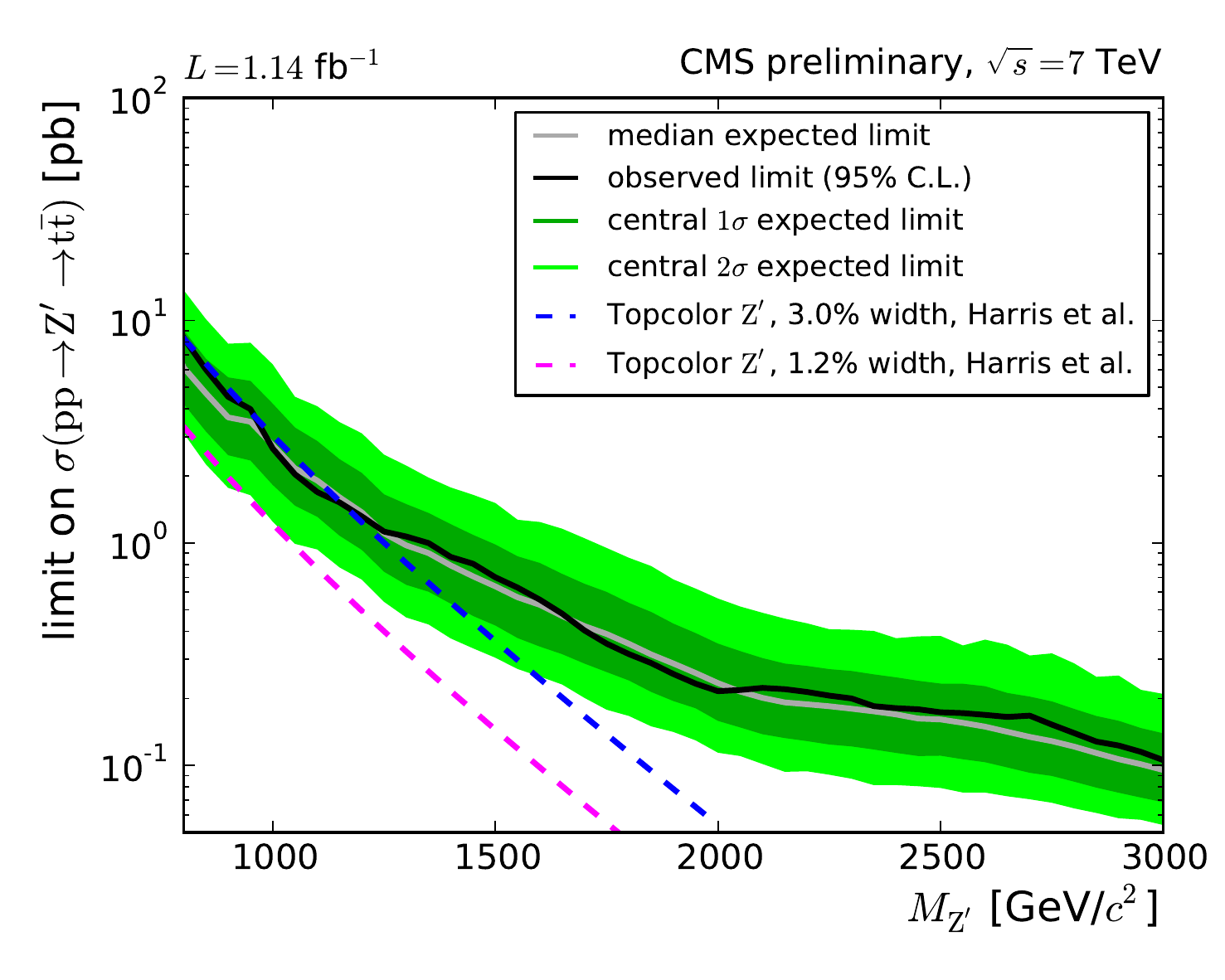} }
    \caption{
    Expected and observed $95\%$ CL upper limits on 
$\sigma(pp\to Z' \to \ttbar)$ for narrow resonances $Z'$, as a function of invariant mass. The topcolor $Z'$ cross section is from \cite{topcolor}, updated to $\sqrt{s} =7 \TeV$ via private communication.
    }
\label{fig:limit_ttsemileptonic}      
\end{figure}

Fig.~\ref{fig:limit_ttsemileptonic} shows the resulting $95\%$ CL upper limits on the cross section for a benchmark topcolor $Z'$ \cite{ttsemileptonic}, computing using a Bayesian method.
 The largest uncertainties come from the jet energy resolution ($10-20\%$) and the jet energy scale ($2-3\%$).   With 1.1\fbinv, 
we exclude a topcolor $Z'$ of width $3\%$ in the mass regions  $805<m_{Z'}<935\GeV$ and $960<m_{Z'}<1060\GeV$.

\section{$t\bar{t}$ Resonances: All-Hadronic Decay}\label{sec:tthadronic}

The motivations for studying the fully hadronic decay of \ttbar are similar to those for the semileptonic search.  Likewise, the all-hadronic search exploits the highly boosted nature of the top quarks from high mass resonances.  Moreover, the all-hadronic decay benefits from a higher branching ratio than the semileptonic decay.

\begin{figure}
\resizebox{1.1\columnwidth}{!}{%
  \includegraphics{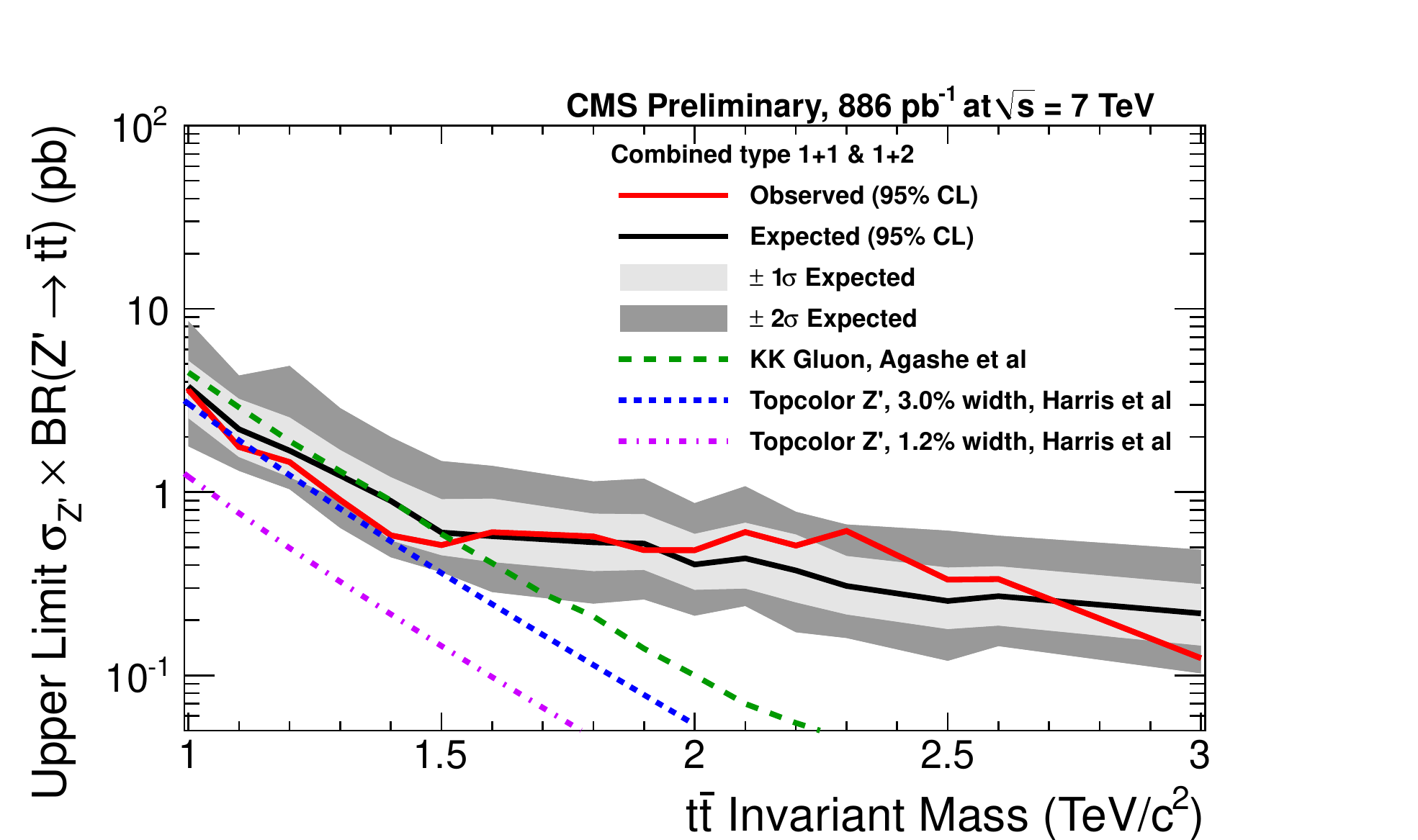} }
\caption{The 95\% CL upper limit on a product of the production cross section of $Z'$ and the branching fraction for its decay into \ttbar pairs, as a function of assumed $Z'$ mass, for a combination of ``1+2" and ``1+1'" channels.   Three theoretical models are examined in the dashed lines: a Kaluza-Klein gluon model and a topcolor $Z'$ model (updated to $\sqrt{s} =7 \TeV$ via private communication)~\cite{topcolor}  with widths $3\%$ and $1.2\%$.
}
\label{fig:limit_tthadronic}      
\end{figure}

In this analysis, each event is divided into hemispheres, such that each hemisphere contains the final products of each top.  Then, the top decays are classified into categories, depending on the how boosted the top is:  (1) ``high boost" tops  are those in which all three jets are merged into one top jet and (2) ``moderate boost" tops are those in which only two out of three of the jets are merged.  We conduct the search in two categories: ``type 1+1", which have two highly boosted top jets, or ``type 1+2", which are three-jet events.  Jets are reconstructed using particle flow and Cambridge-Aachen clustering algorithms.
The dominant background comes from QCD multijets, which is estimated with  a data-driven top-tagging mistag rate. The small continuum \ttbar contribution is estimated with simulation.
The limits are evaluated with a counting experiment, using a Bayesian procedure.
Fig.~\ref{fig:limit_tthadronic} depicts the $95\%$ CL upper limits on the product of the cross section of $Z'$ and the branching ratio for its decay into \ttbar pairs \cite{tthadronic}.
With 886\pbinv, we exclude the KK gluon masses between $1.0-1.5\TeV$.

\section{Conclusions}

We present searches for high mass resonances with the CMS detector in the diphoton, dilepton, dijet, and \ttbar channels.  Observing no excess above standard model predictions, we set limits on a variety of benchmark models, including those predicting gravitons and $Z'$.

\section{Acknowledgements}

We wish to congratulate our colleagues in the CERN accelerator departments for the excellent performance of the LHC machine. We thank the technical and administrative staff at CERN and other CMS institutes, and acknowledge support from: FMSR (Austria); FNRS and FWO (Belgium); CNPq, CAPES, FAPERJ, and FAPESP (Brazil); MES (Bulgaria); CERN; CAS, MoST, and NSFC (China); COLCIENCIAS (Colombia); MSES (Croatia); RPF (Cyprus); MoER, SF0690030s09 and ERDF (Estonia); Academy of Finland, MEC, and HIP (Finland); CEA and CNRS/IN2P3 (France); BMBF, DFG, and HGF (Germany); GSRT (Greece); OTKA and NKTH (Hungary); DAE and DST (India); IPM (Iran); SFI (Ireland); INFN (Italy); NRF and WCU (Korea); LAS (Lithuania); CINVESTAV, CONACYT, SEP, and UASLP-FAI (Mexico); MSI (New Zealand); PAEC (Pakistan); MSHE and NSC (Poland); FCT (Portugal); JINR (Armenia, Belarus, Georgia, Ukraine, Uzbekistan); MON, RosAtom, RAS and RFBR (Russia); MSTD (Serbia); MICINN and CPAN (Spain); Swiss Funding Agencies (Switzerland); NSC (Taipei); TUBITAK and TAEK (Turkey); STFC (United Kingdom); DOE and NSF (USA).



\begin{thebibliography}{}


\bibitem{ref_rsg}
  L.~Randall and R.~Sundrum,
  ``An alternative to compactification,''
  Phys.\ Rev.\ Lett.\  {\bf 83} (1999) 4690.
    
    
\bibitem{ref_gauge}
  E.~Eichten, I.~Hinchliffe, K.~D.~Lane and C.~Quigg,
  ``Super Collider Physics,''
  Rev.\ Mod.\ Phys.\  {\bf 56} (1984) 579.
    
\bibitem{JINST} CMS Collaboration, ``The CMS experiment at the CERN LHC," JINST {\bf 3} (2008) S08004.

\bibitem{diphoton}
CMS Collaboration,
``Search for signatures of extra dimensions in the diphoton mass spectrum at the Large Hadron Collider,"  arXiv:1112.0688 [hep-ex].


\bibitem{CLs1}
  A.~L.~Read,
  ``Presentation of search results: The CL(s) technique,''
  J.\ Phys.\ G {\bf 28}, 2693 (2002).
 
 \bibitem{CLs2}
 T.~Junk,
  ``Confidence Level Computation for Combining Searches with Small Statistics,''
  Nucl.\ Instrum.\ Meth.\  A {\bf 434}, 435 (1999).
  


\bibitem{dilepton}
CMS Collaboration,
``Search for Resonances in the Dilepton Mass Distribution in pp Collisions at $\sqrt{s}$ = 7 TeV,"
CDS Record  {\bf 1369192} (2011).  

   \bibitem{ref_string} 
  L.~A.~Anchordoqui \etal,
 ``Dijet signals for low mass strings at the LHC,''
  Phys.\ Rev.\ Lett.\  {\bf 101} (2008) 241803.
  
\bibitem{ref_diquark}
  J.~L.~Hewett and T.~G.~Rizzo,
 ``Low-Energy Phenomenology of Superstring Inspired E(6) Models,''
  Phys.\ Rept.\  {\bf 183} (1989) 193.

\bibitem{ref_qstar}
  U.~Baur, I.~Hinchliffe and D.~Zeppenfeld,
  ``Excited Quark Production At Hadron Colliders,''
  Int.\ J.\ Mod.\ Phys.\  A {\bf 2} (1987) 1285.

\bibitem{ref_axi}
  P.~H.~Frampton and S.~L.~Glashow,
  ``Chiral Color: An Alternative to the Standard Model,''
  Phys.\ Lett.\  B {\bf 190} (1987) 157.

\bibitem{ref_coloron} 
  E.~H.~Simmons,
 ``Coloron phenomenology,''
  Phys.\ Rev.\  D {\bf 55} (1997) 1678.
  

\bibitem{pflow}
CMS Collaboration,
"Commissioning of the Particle-Flow reconstruction in Minimum-Bias and Jet Events from pp Collisions at 7 TeV,"
CDS Record  {\bf 1279341} (2010).  


\bibitem{dijet}
CMS Collaboration,
``Search for Resonances in the Dijet Mass Spectrum from 7 TeV pp Collisions at CMS,"
Phys.\ Lett.\  B {\bf 704} (2011) 123.






\bibitem{rs_gluon_1}
  K.~Agashe \etal,
  ``LHC signals from warped extra dimensions,''
  Phys.\ Rev.\  D {\bf 77}, 015003 (2008).

\bibitem{ttsemileptonic}
CMS Collaboration,
``Search for heavy narrow resonances decaying to ttbar in the muon+jets channel."
CDS Record {\bf 1376673} (2011).

\bibitem{topcolor}
  R.~M.~Harris, C.~T.~Hill and S.~J.~Parke,
  ``Cross section for topcolor $Z'(t)$ decaying to \ttbar,''
  arXiv:hep-ph/9911288.
  


\bibitem{tthadronic}
CMS Collaboration,
``Search for BSM \ttbar Production in the Boosted All-Hadronic Final State,"
CDS Record  {\bf 1370237} (2011).
  
  
\end{thebibliography}
\end{document}